\documentclass[12pt,a4paper,english]{article}
\usepackage{titling}
\usepackage[affil-it]{authblk}
\usepackage[dvipsnames]{xcolor}
\usepackage{longtable}
\usepackage{lscape}
\usepackage{graphicx}
\usepackage{epsfig}
\usepackage{dsfont}
\usepackage{amsfonts}
\usepackage{amsmath}
\usepackage{amsthm}
\usepackage{amssymb}
\usepackage{bbold}
\usepackage[english]{babel}
\usepackage[utf8]{inputenc}
\usepackage{calc}
\usepackage{cancel}
\usepackage{float}
\usepackage{slashed}
\usepackage{mathrsfs}
\usepackage{pdfpages}
\usepackage{tabu}
\usepackage{simplewick}
\usepackage[hidelinks]{hyperref}
\usepackage{multirow}
\usepackage{hhline}
\usepackage{float}
\usepackage{bm}
\usepackage{braket}
\usepackage{comment}
\usepackage{cite}
\usepackage[left=0.8in,right=0.8in,top=1.0in,bottom=1.2in]{geometry}

\usepackage{datetime}
\usepackage{booktabs}
\date{}

\numberwithin{equation}{section}
\numberwithin{figure}{section}
\numberwithin{table}{section}
 
\newcommand{\be} {\begin{equation}}
\newcommand{\ee} {\end{equation}}
\newcommand{\bea} {\begin{eqnarray}}
\newcommand{\eea} {\end{eqnarray}}
\newcommand{\no} {\nonumber}
\newcommand{\cO}{{\mathcal Q}}

\newcommand{\cL}{{\mathcal L}}
\newcommand{\cH}{{\mathcal H}}
\newcommand{\tcH}{\widetilde{\mathcal H}}

\newcommand{\cA}{{\mathcal A}}
\newcommand{\cF}{{\mathcal F}}
 
\newcommand{\cQ}{{\mathcal Q}} 
\newcommand{\cM}{{\mathcal M}} 
\newcommand{\cN}{{\mathcal N}} 
\newcommand{\llpair}{\ell^+\ell^-}
\newcommand{\mmpair}{\mu^+\mu^-}

\renewcommand{\Im}{\text{Im}}

\def\eq#1{{Eq.~(\ref{#1})}}

\def\Table#1{{Table~\ref{#1}}}

\makeatletter
\g@addto@macro\bfseries{\boldmath}
\makeatother

\allowdisplaybreaks

\newcommand{\published}[1]{%
\gdef\puB{#1}}
\newcommand{\puB}{}

\begin{document}

\title{\textbf{Short- vs. long-distance 
physics  \\ in $B\to K^{(*)} \llpair$: 
a data-driven analysis.}}

\author[1]{Marzia Bordone\thanks{marzia.bordone@cern.ch}}
\author[2]{Gino Isidori\thanks{isidori@physik.uzh.ch}}
\author[2]{Sandro M\"{a}chler \thanks{sandro.maechler@physik.uzh.ch}}
\author[2]{Arianna Tinari \thanks{arianna.tinari@physik.uzh.ch}}

\affil[1]{Theoretical Physics Department, CERN, 1211 Geneva 23, Switzerland}
\affil[2]{Physik-Institut, Universit\"at Z\"urich, CH-8057 Z\"urich, Switzerland}

\published{\flushright CERN-TH-2024-017\vskip2cm}
 
\maketitle 

\begin{abstract}
We analyze data on $B\to K\mmpair$ and  $B\to K^*\mmpair$ decays in the whole dilepton invariant mass spectrum with the aim of disentangling short- vs. long-distance contributions. 
The sizable long-distance amplitudes from $c \overline{c}$ narrow resonances are taken into account by employing a dispersive approach.
For each available  $q^2=m^2_{\mu\mu}$ bin and each helicity amplitude an independent determination of the Wilson coefficient 
$C_9$, describing $b\to s\llpair$ transitions at short distances,  
is obtained. The consistency of the $C_9$ values thus obtained provides an {\em a posteriori} check of the absence of additional, 
sizable, long-distance contributions. The systematic difference of these values from the Standard Model expectation supports the hypothesis of a non-standard $b\to s \mmpair$ amplitude of short-distance origin. 
\end{abstract}

\newpage 

%\tableofcontents 

\section{Introduction}

Exclusive and inclusive $b\to s\llpair$ decays are sensitive 
probes of physics beyond the Standard Model (SM). 
The flavor-changing neutral-current (FCNC) 
structure implies a strong suppression of the decay amplitudes 
within the SM and, correspondingly, enhanced sensitivity to short-distance physics. The two ingredients to fully exploit this potential are precise measurements to be compared with precise theoretical SM predictions. 

On the experimental side, the exclusive $B\to K\mmpair$ and  $B\to K^*\mmpair$ decays are very promising. The LHCb collaboration has already been able to identify large samples of events on both modes with an excellent signal/background ratio, providing precise information on the decay distributions at a fully differential level~\cite{LHCb:2014iah,LHCb:2013vga,LHCb:2013odx}. 
In the $B\to K\mmpair$ case precise results have also been recently reported by the CMS experiment~\cite{CMS-PAS-BPH-22-005}. In all cases the present experimental errors are statistically dominated and are expected to improve significantly in the near future. 

The difficulty of performing precise SM tests via these exclusive modes lies more on the theoretical side, given their theoretical description requires non-perturbative inputs. The latter can be divided into two main categories: i)~the $B\to K^{(*)}$ form factors, necessary to estimate the hadronic matrix elements of the local $b\to s$ operators; ii)~the non-local hadronic matrix elements of four-quark operators related to charm re-scattering. 
While the theoretical error related to the first category can be systematically improved and controlled via Lattice QCD, a systematic tool to deal with the second category, in the whole kinematical region, is not yet available. 

From the observed values of $\Gamma(B\to K^{(*)} J/\Psi)$ and 
$\Gamma(B\to K^{(*)} \Psi(2S))$  we know that charm re-scattering completely obscures the rare FCNC transitions when the invariant mass of the dilepton pair, $q^2 = (p_{\ell^+} + p_{\ell^-})^2$, is in the region of the narrow 
charmonium resonances. This is why precise SM tests in the rare modes are usually confined to the so-called low-$q^2$ ($q^2 \lesssim 6~$GeV$^2$)  and  high-$q^2$ ($q^2 \gtrsim 15\,$GeV$^2$) regions. 
Although, as pointed out in~\cite{Brass:2016efg}, also the 
high-$q^2$ region can be used to 
extract short-distance information 
via a data-driven treatment of 
the resonance contributions.

Estimates of the non-local hadronic matrix elements, 
obtained by combining dispersive methods and 
heavy-quark expansion~\cite{Bobeth:2017vxj,Khodjamirian:2012rm,Khodjamirian:2010vf}, indicate that charm re-scattering is well described by the (small) perturbative contribution in the low-$q^2$ region. Using 
these results, but also with more conservative estimates of charm re-scattering 
(see in particular \cite{Arbey:2018ics}), 
several groups observed a significant tension between data and SM predictions 
in the low-$q^2$ region (see~\cite{Alguero:2023jeh,Alguero:2018nvb,Gubernari:2020eft,Gubernari:2022hxn,Altmannshofer:2021qrr,Hurth:2020ehu,Wen:2023pfq,SinghChundawat:2022zdf, SinghChundawat:2022ldm, LHCb:2023gpo,LHCb:2023gel} for recent analyses).
On the other hand, doubts about a possible underestimate of the theory errors in some of these analyses, particularly those based on dispersive methods,   
have been raised in Ref.~\cite{Ciuchini:2019usw,Ciuchini:2022wbq}.  
 An independent indication consistent with the tension observed at low-$q^2$, despite with lower statistical significance, has been obtained recently in~\cite{Isidori:2023unk} by looking at the semi-inclusive rate in the high-$q^2$ region. 

The purpose of this paper is to extract additional information from data that can shed light on the origin of this tension. More precisely, we extract differential properties on the whole $q^2$ spectrum
about the difference between data and theory that can help distinguish a non-local amplitude (of SM origin) vs.~a short-distance one (of non-SM origin). To achieve this goal,  we put together all the known ingredients of $B\to K^{(*)}\mmpair$ amplitudes within the SM, taking into account also the contribution of charmonium resonances. The latter are described by means of data via a subtracted dispersion relation. We then compare this amplitude with the experimental results for the two rare modes. By doing so, we determine the residual amplitude sensitive to charm re-scattering, both as a function of $q^2$ and as a function of the specific hadronic transition. As we shall show, the residual amplitude extracted in this way shows no significant dependence on $q^2$, nor a dependence on the hadronic transition, contrary to what would be expected from a long-distance contribution. 
These results confirm a similar conclusion obtained first in Ref.~\cite{Alguero:2019ptt}, although with a larger $q^2$ binning, a wider cut to avoid the charmonium-resonance region, and averaging over the different hadronic channels.

The paper is organized as follows. In Section~\ref{sect:theory} we discuss the 
structure of $B\to K^{(*)}\llpair$ amplitudes within the SM, focusing in particular 
on the non-perturbative effects which can mimic the contribution 
of the short-distance operator $\cO_9$. We present both a general parametrization 
of these effects, and an estimate based on dispersion relations. 
In Section~\ref{sect:data} we analyze the available experimental 
data using the amplitude decomposition presented in Section~\ref{sect:theory}, which contains all the known ingredients 
of $B\to K^{(*)}\mmpair$ transitions within the SM, but  is general enough to describe possible additional non-local contributions. The outcome of the 
data-theory comparison is a series of independent determinations of the Wilson coefficient $C_9$ in each $q^2$ bin and each independent hadronic amplitude.
The implications of these results are discussed in Section~\ref{sect:disc} and  
summarized in the Conclusions. The Appendix is devoted to the determination of the parameters appearing in the dispersive description of charmonium resonances.

\section{Theoretical framework}
\label{sect:theory}
The effective Lagrangian describing  $b\to s\ell^+\ell^-$ transitions, after integrating 
out the SM degrees of freedom above the 
$b$-quark mass,
can be written as 
    \begin{equation}
       \cL_\mathrm{eff}(b\to s\ell^+\ell^-) = 
        \frac{4 G_F}{\sqrt{2}}  V_{tb}V^*_{ts}
  \sum_{i=1}^{10} C_i \cQ_i  ~ + ~\cL_{\rm QCD+QED}^{[N_f=5]} \,.
    \end{equation}
    Here $V_{ij}$ denote the elements of the Cabibbo-Kobayashi-Maskawa (CKM) matrix,
    and the subleading terms proportional to  
    $V_{ub}V^*_{us}$ have been neglected (i.e.~we assume $V_{cb}V^*_{cs} \approx - V_{tb}V^*_{ts}$).
    The most  relevant effective operators are 
    \begin{align}
        \cQ_1 =& (\bar{s}^\alpha_{L}\gamma_\mu c^\beta_L)(\bar{c}^\beta_L\gamma^\mu b^\alpha_L)\,,   &  \cQ_2 =&(\bar{s}_{L}\gamma_\mu c_L)(\bar{c}_L\gamma^\mu b_L)\,,   \\
        \cQ_7=&\frac{e}{16\pi^2}m_b(\bar{s}_{L}\sigma^{\mu\nu} b_R)F_{\mu\nu}\,,    & \cQ_8=&\frac{g_s}{16\pi^2}m_b(\bar{s}_{L}\sigma^{\mu\nu}T^a b_R)G_{\mu\nu}^a\,,   \\
        \cQ_9=&\frac{e^2}{16\pi^2}(\bar{s}_{L}\gamma_\mu b_L)(\bar{\ell}\gamma^\mu \ell)\,,    &  \cQ_{10}=&\frac{e^2}{16\pi^2}(\bar{s}_{L}\gamma_\mu b_L)(\bar{\ell}\gamma^\mu\gamma_5 \ell)\,.
    \end{align}
    The explicit form of the additional four-quark operators
    $\cQ_{3-6}$, with 
    subleading Wilson coefficients, can be found in~\cite{Altmannshofer:2008dz}.
  
    Only the FCNC quark bilinears 
    $\cQ_{7,9,10}$ have 
    non-vanishing tree-level matrix elements in $B\to K^{(*)}\mmpair$. Those of the 
    operators  $\cQ_{7}$ and  $\cQ_{9}$, which are central to our analysis, lead to the following contributions to the decay amplitudes
   \begin{eqnarray}
&&    \left.\cM\left(B\rightarrow K\ell^+ \ell^-\right)\right\vert_{C_{7,9}}  =
      2 \cN \left[
        C_9 \langle K | \bar{s}_{L}\gamma_\mu b_L | B \rangle 
        -\frac{2 m_b}{q^2}
        C_7 \langle K | \bar{s}_{L} i \sigma_{\mu\nu} 
        q^\nu b_R | B \rangle \right]  \ell \gamma^\mu \ell  \no\\
     && \qquad  =  \cN\, C_9\, \Bigg[ 
         f_+(q^2) (p_B+p_K)^\mu + f_-(q^2)q^\mu  \Bigg] \ell \gamma^\mu \ell \no\\
     && \qquad 
        +\,  \cN\, C_7\,  \frac{ f_T(q^2)}{(m_B+m_K)}
        \Bigg[ q^2 (p_B+p_K)^\mu - (m_B^2-m_K^2) q^\mu \Bigg]   \left(\frac{2 m_b}{q^2}\right) \ell \gamma^\mu \ell
        \label{eq:BToHI0} 
   \end{eqnarray} 
and
   \begin{eqnarray}        
&&  \left.\cM\left(B\rightarrow K^*\ell^+ \ell^- \right)\right\vert_{C_{7,9}}  = 
             \cN\, C_9\,   
             \Bigg[ -   2 i \epsilon_{\mu\nu\rho\sigma} (\epsilon^{\ast})^\nu p_B^\rho p_{K^*}^\sigma 
            \frac{  V(q^2) }{m_B+m_{K^*}}  \no \\
     && \qquad\qquad\quad     
             + q_\mu \left(\epsilon^{\ast} \cdot q \right) \frac{2m_{K^*}}{q^2} A_0(q^2)        + \left(\epsilon_{\mu}^{\ast} 
            -q_\mu \frac{\epsilon^{\ast}\cdot q }{q^2}\right) 
            \left(m_B + m_{K^*} \right) A_1(q^2) \nonumber \\
     &&  \qquad\qquad\quad   - \left(\left(p_B+p_{K^*} \right)_\mu -q_\mu \frac{m_B^2-m_{K^*}^2}{q^2} \right) 
            \frac{\epsilon^{\ast} \cdot q}{ m_B+m_{K^*} } A_2(q^2) 
            \Bigg] \bar \ell \gamma^\mu \ell \no  \\
    &&  \quad\   + \cN\, C_7\,   
             \Bigg[ - 2 i \epsilon_{\mu\nu\rho\sigma} (\epsilon^{\ast})^\nu p_B^\rho p_{K^*}^\sigma T_1(q^2)
             +  \left(\epsilon^{\ast} \cdot q \right) 
          \left(q_\mu - \frac{q^2}{m_B^2- m_{K^*}^2} \left(p_B+p_{K^*} \right)_\mu \right) T_3(q^2) 
               \nonumber \\
      &&  \qquad \qquad\quad 
          +\left(\epsilon_{\mu}^{\ast} ( m_B^2- m_{K^*}^2) - 
          \left(\epsilon^{\ast} \cdot q \right)\left(p_B+p_{K^*} \right)_\mu \right) T_2(q^2) \Bigg]
      \left(\frac{2 m_b}{q^2}\right) \ell \gamma^\mu \ell\,,
            \label{eq:BToHII0} 
\end{eqnarray}
where
\begin{equation}
q^\mu=p_B^\mu-p_{K^{(*)}}^\mu\,, \qquad 
\cN=  \sqrt{2} G_\mathrm{F} \alpha_{\rm em} V_{tb} V_{ts}^\ast/(4\pi)\,, 
\end{equation}
while  
$\{A_i(q^2), V(q^2), T_i(q^2)\}$ 
and $f_{\pm,T}(q^2)$ denote the relevant
$B\to K^*$ and $B\to K$ hadronic local form factors, respectively. 
Due to the conservation of the leptonic current ($q_\mu \ell \gamma^\mu \ell=0$), 
we can rewrite Eqs.~(\ref{eq:BToHI0})--(\ref{eq:BToHII0})  as
\begin{eqnarray}
    &&  \left.\cM\left(B\rightarrow K\ell^+ \ell^-\right)\right\vert_{C_{7,9}} =
        \cN\,  \left[ C_9 +\frac{2 m_b }{ m_B +m_K}         
        \frac{f_T(q^2)}{f_+(q^2) }  C_7 \right] \,  
        f_+(q^2) (p_B +p_K)_\mu\,  \bar \ell \gamma^\mu \ell \no\\
    &&  \left.\cM\left(B\rightarrow K^*\ell^+ \ell^- \right)\right\vert_{C_{7,9}} 
        = \cN\,   \Bigg\{  \no \\
    &&  \quad    
            - \left[ C_9 + 
            \frac{ 2 m_b (m_B+m_{K^*} )}{ q^2} 
            \frac{ T_1(q^2)}{V(q^2)} C_7 \right]
            \frac{   2 V(q^2) }{m_B+m_{K^*}} 
            i \epsilon_{\mu\nu\rho\sigma} (\epsilon^{\ast})^\nu p_B^\rho p_{K^*}^\sigma  
            \no\\
    &&      \quad -\left[C_9  
            + \frac{2 m_b (m_B+m_{K^*})}{q^2} \frac{T_2(q^2)}{A_2(q^2)} C_7 \left(1+ O\Big(\frac{q^2}{m_B^2}\Big)\right) \right]
            \frac{  A_2(q^2) }{ m_B+m_{K^*} }     (\epsilon^{\ast} \cdot q)(p_B+p_{K^*})_\mu 
            \no\\
    &&      \quad +\left[C_9 
            + \frac{ 2 m_b(m_B^2- m_{K^*})}{q^2} \frac{T_2(q^2)}{A_1(q^2)} C_7 \right] A_1(q^2) \left(m_B + m_{K^*} \right) \epsilon_{\mu}^{\ast} 
            \Bigg\}\, \bar \ell \gamma^\mu \ell\,,
            \label{eq:C9matrixel}    
\end{eqnarray}
where it emerges more clearly that the contributions of  $\cQ_{7}$ and  $\cQ_{9}$
admit the same Lorentz decomposition. The four independent Lorentz structures
appearing in these amplitudes are in a linear relation with the four independent  $|B\rangle \to |H_\lambda \rangle$ hadronic transitions, where
\begin{equation}
| H_K \rangle \equiv  | K \rangle\,, \qquad   |H_\perp \rangle \equiv  | K^*(\epsilon_\perp) \rangle\,, \qquad   
| H_\parallel \rangle \equiv  | K^*(\epsilon_\parallel ) \rangle\,, \qquad   
| H_0 \rangle \equiv  | K^*(\epsilon_0 ) \rangle\,.
 \label{eq:Hlamb}
\end{equation}
To identify these amplitudes rather than the  Lorentz structures,
we redefine the independent form-factor combinations
associated to the matrix element of the $\cO_9$ operator as  
\begin{align}
 \cF_K(q^2)&=  f_+(q^2) \,, \qquad\quad  
  \cF_\perp(q^2) = V(q^2) \,,  \qquad\quad  
 \cF_\parallel(q^2) =  A_1(q^2) \,,  
  \nonumber \\ 
  \cF_0(q^2)  &=  \frac{ (m_B+m_{K^*})^2( m_B^2-m_{K^*}^2-q^2) A_1(q^2)
- \lambda(m_B^2,m_{K^*}^2, q^2) A_2(q^2)}{16 m_B m_{K^*}^2 (m_B+m_{K^*})}\,,
\label{eq:helicities}
\end{align}
with $\lambda(a,b,c)=a^2+b^2+c^2-2(ab+ac+bc)$.

\subsection{Hadronic matrix elements of four-quark operators}

The focus of this paper is to extract information on the non-local matrix elements of the four-quark operators $\cQ_{1,6}$ from data.
    To this purpose, the first point to note is that to all orders in $\alpha_s$, and to first order in $\alpha_{\rm em}$, 
    these matrix elements have the same structure as the matrix elements of $\cQ_7$
    and $\cQ_9$. In other words, Lorentz and gauge invariance 
    imply  
\begin{align}
 & \left.\cM\left(B\rightarrow K\ell^+ \ell^-\right)\right\vert_{C_{1-6}}  =\ - i \frac{ 32 \pi^2\cN}{q^2}\,  \bar \ell \gamma^\mu \ell  
\int d^4 x e^{iqx} \langle H_\lambda   | T\left\{     
j_{\rm \mu}^{\rm em}(x), \sum_{i=1,6} C_i\cQ_i (0)
\right\} | B  \rangle \nonumber \\
& \qquad\qquad  = 
 \left( \Delta^\lambda_9(q^2) + \frac{m_B^2}{q^2} \Delta^\lambda_7
\right) \langle H_\lambda\ \ell^+ \ell^- |  \cQ_9  | B  \rangle \,,
\label{eq:MQi}
\end{align}
where  $j_\mu^\text{em} (x)$ denotes the electromagnetic current,
\begin{equation}
j_\mu^\text{em}(x)= \sum_{q=u,c} \frac{2}{3}\, (\bar{q}\gamma_\mu q)(x) -\sum_{q=d,s,b} \frac{1}{3}\, (\bar{q}\gamma_\mu q) (x)\,,
\end{equation}
and the explicit form of $\langle H_\lambda\ \ell^+ \ell^- |  \cQ_9  | B  \rangle$ follows from Eq.~(\ref{eq:C9matrixel}). 

The function $\Delta^\lambda_9(q^2)$ parameterizes the matrix elements of the four-quark operators in all the kinematical range
but for possible singular contributions in the $q^2\to 0$ limit.
By construction, $\Delta^\lambda_9(q^2)$  is a function of $q^2$ and, a priori, differs for each hadronic amplitude (i.e.~for each value of $\lambda\in\{K,\perp, \parallel, 0\}$). The 
coefficient $\Delta^\lambda_7$ describes 
singular contributions in $q^2\to 0$ limit, 
corresponding to matrix elements of the four-quark operators which are non-zero 
for $B \to H_\lambda \gamma$ with an on-shell photon.
By helicity conservation, this is possible only $\lambda=\perp$ and $\parallel$, hence  $\Delta^{K,0}_7=0$.

In principle, we should introduce  additional independent (non-local) hadronic form factors to describe the matrix elements of the  four-quark operators; however, thanks to Eq.~(\ref{eq:MQi}), we can effectively describe these matrix elements via  
an appropriate $q^2$-- and $\lambda$--dependent modifications of  $C_9$ and a universal ($q^2$-- and $\lambda$--independent) shift of $C_7$.
In particular, the regular terms in the $q^2\to 0$ limit
are described by Eq.~\ref{eq:C9matrixel}
via the substitution 
\begin{equation}
    C_9 \to C_9^{\mathrm{eff}, \lambda} = C_9 + Y^\lambda (q^2)~.
    \label{eq:C9eff}
\end{equation}

By means of Eq.~(\ref{eq:C9eff})  we describe in full generality both perturbative and non-perturbative contributions.
The perturbative ones, evaluated at the lowest-order in $\alpha_s$, lead to the well-known  hadronic-independent (factorizable) expression
\begin{equation}
\left.    Y^\lambda (q^2) \right|_{\alpha_s^0}  =  Y^{[0]}_{\rm q\bar q} (q^2) + Y^{[0]}_{\rm c\bar{c}} (q^2) +Y^{[0]}_{\rm b\bar b} (q^2)\,,
\label{eq:Ypert}
\end{equation}
where
\begin{align*}
     Y^{[0]}_{\rm q\bar q} (q^2) &= \frac{4}{3}C_3 + \frac{64}{9} C_5 + \frac{64}{27} C_6 - \frac{1}{2} h(q^2, 0) \left( C_3 +\frac{4}{3} C_4 + 16 C_5 + \frac{64}{3} C_6 \right)\,, \\
   Y^{[0]}_{\rm c\bar c} (q^2)  &= h(q^2, m_c) \left( \frac{4}{3} C_1 + C_2 + 6 C_3 + 60 C_5 \right)\,,  \\
  Y^{[0]}_{\rm b\bar b} (q^2)    &= - \frac{1}{2} h(q^2, m_b) \left( 7 C_3 + \frac{4}{3} C_4 + 76 C_5 + \frac{64}{3} C_6 \right)\,,
\end{align*}
with
\begin{equation*}
    h(q^2, m) = -\frac{4}{9} \left( \ln{\frac{m^2}{\mu^2}} - \frac{2}{3} - x \right) - \frac{4}{9} (2+x) \begin{cases}
        \sqrt{x-1} \arctan{\frac{1}{\sqrt{x-1}}} \,,  & x=\frac{4 m^2}{q^2} >1\,,
        \\
        \sqrt{1-x} \left(\ln{\frac{1+\sqrt{1-x}}{\sqrt{x}}} - \frac{i \pi}{2}\right), & x=\frac{4 m^2}{q^2} \leq 1\,.
    \end{cases}
\end{equation*}
To a good accuracy, given the smallness of $C_{3 \ldots 6}$, it follows that 
\begin{equation}
\left.    Y^\lambda (q^2) \right|_{\alpha_s^0}  \approx   h(q^2, m_c) \left( \frac{4}{3} C_1 + C_2 \right)\,.
\label{eq:Ypert2}
\end{equation}
Non-factorizable corrections to $Y^\lambda(q^2)$
are generated at higher order in $\alpha_s$. 
We checked against the \texttt{EOS} software~\cite{EOSAuthors:2021xpv} that
the perturbative non-factorizable corrections to $C_9^{\rm eff}$
 are numerically subleading and can be safely neglected.\footnote{~We thank M\'{e}ril Reboud for providing the necessary information to perform the cross-checks.} This is not the case for the non-perturbative contributions induced by 
$c \overline{c} $ resonances, which we discuss in detail in Section~\ref{eq:nonpertcc}.

The contributions of the four-quark operators leading to a non-vanishing $B \to H_\lambda \gamma$ amplitude, hence 
generating a pole for $q^2\to 0$
in $B \to H_\lambda \ell^+ \ell^-$,
have been analyzed first in Ref.~\cite{Beneke:2001at}.
Here there are no factorizable contributions. 
The leading effect has been estimated in perturbation theory up to the next-to-leading order  in $\alpha_s$
in the heavy quark limit.
As anticipated, a pole for $q^2\to 0$ occurs 
only in the $\lambda=\perp$ and $\parallel$ amplitudes, whose tensor form factors coincide in the $q^2 \to 0$ limit. This is why these contributions can effectively be taken into account by a universal shift of 
$C_7$~\cite{Beneke:2001at}:
\begin{equation}
C_7 \to C_7^\mathrm{eff} \approx 1.33 C_7\,.
\end{equation}
We implement this shift in 
all $C_7$ terms in 
Eq.~(\ref{eq:C9matrixel}).\footnote{~This shift, which has been determined in the $q^2\to 0$ limit, provides a correct evaluation of 
non-factorizable corrections to the pole terms only; however, given the smallness of non-pole terms proportional to $C_7$, we can safely 
apply it as universal shift in all $C_7$ terms.}
In order to take into account the scale uncertainty and missing higher-order corrections, we assign a conservative $10\%$ error to the value of $C_7^\mathrm{eff}$.

\subsection{Long-distance contribution from  \texorpdfstring{$c \overline{c} $}{} resonances}
\label{eq:nonpertcc}
The perturbative result in  Eq.~(\ref{eq:Ypert2}) does not provide a good approximation 
of the large non-perturbative contribution induced by the narrow charmonium resonances.
However, the latter can be well described using dispersion relations and 
experimental data~\cite{Khodjamirian:2012rm,Lyon:2014hpa,Blake:2017fyh,Bobeth:2017vxj,Cornella:2020aoq}. To achieve this goal, we need to go back to  Eq.~(\ref{eq:MQi}) and isolate the hadronic part of the matrix elements.
In the $B\to K$ case, this can can be decomposed as~\cite{Khodjamirian:2012rm}
    \begin{align}
    \begin{aligned}
     - i \int d^4x e^{iqx}\langle K |T\left\{j_\mu^\text{em}(x), \sum_{i=1,2} C_i \cQ_i (0) \right\}| B\rangle 
        =\,[q^2 (p_B)_\mu - (p_B\cdot q)q_\mu]\cH^K_{\rm{c\bar c}}(q^2)\,.
    \end{aligned}
    \end{align}
    Proceeding in a similar manner, 
    we decompose the four-quark 
    matrix elements in $B\to K^*$ as
\begin{align}
  &  - 2i \int d^4x e^{iqx}\langle K^* |T\left\{j_\mu^\text{em}(x), \sum_{i=1,2} C_i \cQ_i (0) \right\}| B\rangle = 
        \nonumber \\  
&   \quad  =  \left(\epsilon_{\mu}^{\ast} 
            -q_\mu \frac{\epsilon^{\ast}\cdot q }{q^2}\right) 
            \left(m_B + m_{K^*} \right) \cH^{\parallel}_{\rm{c\bar c}}(q^2)
            -i \epsilon_{\mu\nu\rho\sigma} (\epsilon^{\ast})^\nu p_B^\rho p_{K^*}^\sigma 
            \frac{2 }{m_B+m_{K^*}} \cH^\perp_{\rm{c\bar c}}(q^2)  
            \nonumber \\
     &\quad    -   \left(\left(p_B+p_{K^*} \right)_\mu -q_\mu \frac{q\cdot (p_B+p_{K^*})}{q^2} \right) 
            \frac{\epsilon^{\ast} \cdot q}{ m_B+m_{K^*} }  \tcH^{0}_{\rm{c\bar c}}(q^2)\,. 
            \label{eq:BKsdec}
\end{align}
Since we are interested in labeling the amplitudes according to the helicity of the hadronic state, in analogy with  
Eq.~(\ref{eq:helicities}), we also define
\begin{equation}
      \cH^0_{\rm{c\bar c}}(q^{2})  =
        \frac{(m_B+m_{K^*})^2 ( m_B^2-m_{K^*}^2-q^2) \cH^\parallel_{\rm{c\bar c}}(q^{2})
        -\lambda(m_B^2,m_{K^*}^2, q^2)   \tcH^0_{\rm{c\bar c}}(q^{2})}{16m_B m_{K^*}^2 (m_B+m_{K^*})}\,.  
\end{equation}

    \begin{table}[t]
            \begin{center}
            \renewcommand{\arraystretch}{1.1} 
            \begin{tabular}{c|c|c||c|c}
                $V$ & $\eta^K_V$ & $\delta^K_V$ &  $m_V$(MeV)  & $\Gamma_V$(MeV) \\
                \hline
                $J/\psi$ & $32.3 \pm 0.6\phantom{0} $ & $-1.50 \pm 0.05 $   & $3096.9$ & $0.0926 \pm 0.0017$ \\
                $\psi(2S)$ & $7.12\pm 0.32 $ & $\phantom{+}2.08 \pm 0.11 $  & $3686.1$ & $0.294 \pm 0.008$\\
                $\psi(3770)$ & $\left(1.3 \pm 0.1 \right) \times 10^{-2}$ & $ -2.89 \pm 0.19 $ & $3773.7 \pm 0.4$   &  $27.2\pm 1.0\ $\\
                $\psi(4040)$ & $\left(4.8 \pm 0.8 \right) \times 10^{-3}$ & $ -2.69 \pm 0.52 $ & $4039 \pm 1$   &  $80\pm 10$\\
                $\psi(4160)$ & $\left(1.5 \pm 0.1 \right) \times 10^{-2}$ & $ -2.13 \pm 0.33 $ & $4191 \pm 5$   &  $70\pm 10$\\
                $\psi(4415)$ & $\left(1.1 \pm 0.2 \right) \times 10^{-2}$ & $ -2.43 \pm 0.43 $ & $4421 \pm 4$   &  $62\pm 20$\\
            \end{tabular}
            \end{center}
            \caption{Magnitudes ($\eta^K_V$) and phases ($\delta^K_V$)
             of the $B^+ \rightarrow K^+ V \to 
             K^+ \mu^+\mu^-$ amplitudes, as determined in Appendix~\ref{app:etaV}.
             The mass and width of the resonances are also reported.
            \label{tab:EtasPhisBtoK}}
        \end{table}

We can write a one-time subtracted dispersion relation for each $\cH^\lambda_{\rm{c\bar c}}(q^{2})$  function, namely 
        \begin{align}
        \begin{aligned}
            \Delta \cH^\lambda_{\rm{q \bar q}}(q^{2})  = \frac{q^2- q_{0}^2}{\pi} \int_{s_0}^\infty ds \frac{ \Im[\cH^\lambda_{\rm{c\bar c}}(s)]}{(s-q_{0}^{2})(s-q^2)}  \equiv \frac{q^2-q_{0}^2}{\pi}\int_{s_0}^\infty ds \frac{\rho^\lambda_{\rm{c\bar c}}(s)}{(s-q_{0}^{2})(s-q^2)}\,.
        \end{aligned}
        \end{align}        
This  allows us to rewrite in full generality (i.e.~without any expansion in $\alpha_s$) the $c\bar c$ contribution to $Y^\lambda (q^2)$ as       
 \begin{align}
Y^\lambda_{\rm{c\bar c}}(q^{2}) =  Y^\lambda_{\rm{c\bar c}}(q^2_0) +  \frac{16 \pi^{2}}{\cF_\lambda (q^2)} \Delta \cH^\lambda_{\rm{c\bar c}}(q^{2})\,,
\label{eq:YdecNP}
 \end{align}           
where $\cF_\lambda (q^2)$ denote the four hadronic form factors defined in Eq.~(\ref{eq:helicities}).

The function $\rho^\lambda_{\rm{c\bar c}}(s)$ is the spectral density for an intermediate hadronic state with $c\bar c$ valence-quark content 
and invariant mass $s$, and  $s_0$ denotes the energy threshold where such state can be created on-shell. The parameter $q_0^2$ is the subtraction point. 
As shown in~\cite{Khodjamirian:2012rm}, one recovers Eq.~(\ref{eq:Ypert2}) if $\rho^\lambda_{\rm{c\bar c}}(s)$ is evaluated at the partonic level,
i.e.~factorizing the hadronic matrix elements as
\begin{align}
 \langle H^\lambda |T\left\{j_\mu^\text{em}(x),  \cO_{1,2} (0) \right\} | B\rangle \propto  \langle 0  |T\left\{j_\mu^\text{em}(x), (\bar{c}_L\gamma^\mu c_L)  (0) \right\} | 0 \rangle \times 
  \langle H^\lambda |  \bar{s}_{L}\gamma_\mu s_L  | B \rangle 
\end{align}
and evaluating the $T$-product between the charm current and $j_\mu^\text{em}(x)$ at $O(\alpha_s^0)$.

In order to take into account non-perturbative effects, we need to evaluate $\rho_{\rm{c \bar c}}(s)$ at the hadronic level.
In this case, the leading contribution is provided by single-particle intermediate states with the correct quantum numbers and valence quarks, namely the 
spin-1 charmonium resonances ($V=J/\psi,  \psi(2S),  \ldots$). Describing these
contributions to $\rho_{\rm{c \bar c}}(s)$ via  a  sum of Breit-Wigner distributions
leads to 
\begin{align} 
            \Delta \cH_{\rm{c\bar c}}^{\lambda,\text{1P} } (q^2) = \left. \sum_{V}\eta^\lambda_V e^{i\delta^\lambda_V}\frac{(q^2-q_0^2)}{(m_V^2-q_0^2)} A_V^\text{res}(q^2)
            \right|_{q^2_0=0} =   \sum_{V }
            \eta^\lambda_V e^{i\delta^\lambda_V} \frac{q^2}{m^2_{V} } A_V^\mathrm{res}(q^2)\,,
          \label{eq:etaVdef}     
\end{align}
where 
\begin{align}
             A_V^\text{res}(q^2)  = \frac{m_V \Gamma_V}{m_V^2-q^2-i m_V \Gamma_V}\,.
\end{align}
The $\{\eta^\lambda_V, \delta^\lambda_V\}$ parameters need to be determined from data. In Table~\ref{tab:EtasPhisBtoK} and Table~\ref{tab:EtasPhisBtoKstar} we report their values for the two leading charmonium resonances, $J/\psi$ and $\psi(2S)$.
In the $B\to K$ case we also report the 
 $\{\eta^\lambda_V, \delta^\lambda_V\}$ for
 the wider charmonium states (which have a smaller impact). The determination of these parameters is discussed in Appendix~\ref{app:etaV}.

        \begin{table}[t]
            \centering
            \begin{tabular}{c|c|c|c}
                $V$ & Polarization & $\eta^\lambda_V$ & $\delta^\lambda_V$ \\
                \hline
                 \multirow{3}{*}{$J/\psi$} & $\perp$ & $26.6 \pm 1.1 $ & $1.46 \pm 0.06 $ \\
                &$\parallel$ & $ 12.3 \pm 0.5 $ & $-4.42 \pm 0.06 $ \\
                & 0 & $13.9 \pm 0.5 $ & $-1.48 \pm 0.05 $ \\
                \hline
                \multirow{3}{*}{$\psi(2S)$} & $\perp$ & $3.0 \pm 0.9 $ & $3.2 \pm 0.4 $ \\
                &$\parallel$ & $1.11 \pm 0.30 $ & $-3.32 \pm 0.22 $ \\
                & 0 & $1.14 \pm 0.06 $ & $2.10 \pm 0.11 $ \\
            \end{tabular}
            \caption{Magnitudes ($\eta^\lambda_V$) and phases ($\delta^\lambda_V$)
             of the $B \rightarrow K^*(\epsilon^\lambda) V \to 
             K^*(\epsilon^\lambda) \mu^+\mu^-$ amplitudes, as determined in Appendix~\ref{app:etaV}.}
            \label{tab:EtasPhisBtoKstar}
        \end{table}

In order to use the general decomposition in Eq.~(\ref{eq:YdecNP}), the last missing ingredient is
the subtraction term $Y^\lambda_{\rm{c\bar c}}(q^2_0)$.
Having chosen as subtraction point $q_0^2=0$, which is far from the resonance region, we can use the perturbative result in Eq.~(\ref{eq:Ypert2}).
Since 
\begin{equation}
    h(q^2, m)\,   \stackrel{q^2 \to 0}{\longrightarrow}\,  - \frac{4}{9} \left[ 1+
    \ln\left( \frac{m^2}{\mu^2} \right) \right]
    \label{eq:hfun}
\end{equation}
we finally obtain         
 \begin{align}
Y^\lambda_{\rm{c\bar c}}(q^{2}) =   - \frac{4}{9} 
    \left[\frac{4}{3} C_1(\mu)  + C_2(\mu) \right] \left[ 1+
    \ln\left( \frac{m^2}{\mu^2} \right) \right]
 +  \frac{16 \pi^{2}}{\cF_\lambda (q^2)}  
   \sum_{V }
            \eta^\lambda_V e^{i\delta^\lambda_V} \frac{q^2}{m^2_{V} } A_V^\mathrm{res}(q^2)\,.
\end{align} 
   
\section{Numerical analysis of the experimental data}
\label{sect:data}

\begin{table}[t]
    \centering
    \begin{tabular}{c|c}
       Parameter & value \\ \hline 
        $\eta_{\mathrm{EW}}G_\mathrm{F}$ & $(1.1745\pm 0.0023) \times 10^{-5} \text{ GeV}^{-2}$ \\
         $ m_c $ & $ 1.68 \pm 0.20 ~ \text{GeV} $ \\
          $ m_b $ & $ 4.87 \pm 0.20 ~ \text{GeV} $ \\
        $ 1/\alpha_{\rm em}(m_b) $ & $ 133  $ \\
        $|V_{tb} V_{ts}^*| $ & $ 0.04185 \pm 0.00093 $ \\ 
            \end{tabular}
            \caption{Input parameter for the numerical analysis.}
            \label{tab:inputsFit}
        \end{table}

\begin{table}[t]
    \centering
    \begin{tabular}{l|c||l|c}
        Coefficient & value  &  Coefficient & value    \\ \hline
                $ C_1(\mu_b) $ & $ -0.291 \pm 0.009 $ &  $ C_6(\mu_b) $ &  $0.0012 \pm 0.0001$ \\
                $ C_2(\mu_b) $ & $ 1.010 \pm 0.001 $ &  $ C_7^{\mathrm{eff}}(\mu_b) $ & $ -0.450\pm 0.050 $ \\
                $ C_3(\mu_b) $ & $ -0.0062 \pm 0.0002 $ & $ C_8^{\mathrm{eff}}(\mu_b) $ & $ -0.1829 \pm 0.0006 $ \\
                $ C_4(\mu_b) $ & $ -0.0873 \pm 0.0010 $ &  $ C_9(\mu_b) $ & $ 4.273 \pm 0.251 $ \\
                $ C_5(\mu_b) $ & $ 0.0004 \pm 0.0010 $ &  $ C_{10}(\mu_b) $ & $ -4.166 \pm 0.033 $ \\
            \end{tabular}
            \caption{Input values for the Wilson Coefficients.}
            \label{tab:inputCi}
        \end{table}

%\subsection{Description of the strategy}

In this section, we describe the fitting procedure that we employ to analyse  the available experimental data on
$B^+\rightarrow K^+ \mmpair$ and 
$B \rightarrow K^{*0} \mmpair$
 differential decay distributions, allowing $C_9$ to 
vary in the most general way.
More precisely, we fit data using 
the SM expressions 
discussed in Section~\ref{sect:theory},
setting 
\begin{equation}
    C_9 \to  C^\lambda_9 (q^2) + Y^\lambda_{\rm{c\bar c}}(q^{2}) +
    Y^{[0]}_{\rm q\bar q} (q^2) +Y^{[0]}_{\rm b\bar b} (q^2)\,,
    \label{eq:C9fit}
\end{equation}
and extracting $C^\lambda_9 (q^2)$ in each  $q^2$ bin and for each value of $\lambda$.
We use the input parameters reported in Table~\ref{tab:inputsFit} and we fix the renormalization scale to $\mu_b = 4.2$ GeV. The SM values of the Wilson coefficients are reported in 
Table~\ref{tab:inputCi}, with errors taking into account the variation of the scale between $\mu_b/2$ and $2 \mu_b$.
 The only case where  the error is larger than what obtained from the na\"ive scale variation is $C_9$: here we adopt
the   estimate presented in Ref.~\cite{Isidori:2023unk}  which conservatively  takes into account also the scale variation associated to the $h$ function in 
Eq.~(\ref{eq:hfun}).\footnote{ The  value of 
 $C_9$ in Table~\ref{tab:inputCi}  does not enter directly our numerical analysis,
 aimed at extracting  $C_9$ from data, 
 but rather provides the reference SM value to compare with the data-driven results.  }

%%%%%%%%%%%%%%%%%%%%%%%%%%%%%%%%%%%%%%%%%%%%%%%%%%%%%%%%%%%%%%%%%%%%%%%%%%%%%%%%
\begin{table}[t]
\begin{minipage}{0.35\textwidth}
\begin{tabular}{ c | c  }
$q^2$ (GeV${^2})$ & $C^K_9$ \\
\midrule
\midrule
$[1.1,2]$ & $1.9_{-0.8}^{+0.5}$\\[2mm]
$[2, 3]$ & $3.2_{-0.4}^{+0.3}$  \\[2mm]
$[3, 4]$ & $2.6_{-0.5}^{+0.4}$   \\[2mm]
$[4, 5]$ &  $2.1_{-0.7}^{+0.5}$  \\[2mm]
$[5, 6]$ &  $2.4_{-0.6}^{+0.4}$   \\[2mm]
$[6, 7]$ &  $2.6_{-0.5}^{+0.4}$  \\[2mm]
$[7,8]$ &   $2.3_{-0.7}^{+0.5}$  \\[2mm]
\midrule
constant & $2.4_{-0.5}^{+0.4} ~ ~ (\chi^2/\mathrm{dof} = 1.35$) \\
\end{tabular}
\end{minipage}
\hfill
\begin{minipage}{0.55\textwidth}
\centering
\begin{tabular}{ c | c | c }
$q^2$ (GeV${^2})$ & $C^K_9$ (LHCb) & $C^K_9$ (CMS) \\
\midrule
\midrule
$[15, 16]$ & $1.8_{-1.8}^{+0.8}$ & $1.4_{-1.4}^{+0.9}$ \\[2mm]
$[16, 17]$ & $2.1_{-1.0}^{+0.7}$ & $1.9_{-1.9}^{+0.8}$  \\[2mm]
$[17, 18]$ & $2.9_{-0.5}^{+0.5} $ & $3.0_{-0.6}^{+0.5}$  \\[2mm]
$[18, 19]$ & $2.7_{-0.5}^{+0.6} $ & \\[2mm]
$ [18,19.24]$ & & $2.9_{-0.7}^{+0.6}$ \\[2mm]
$[19, 20]$ & $0_{-0}^{+1.6} $ &  \\[2mm]
$[20, 21]$ & $1.4_{-1.4}^{+0.9}$&  \\[2mm]
$[21, 22]$ & $3.2_{-0.9}^{+0.8} $& \\[2mm]
$[19.24, 22.9]$ & & $2.5_{-1.0}^{+0.7}$ \\[2mm]
\midrule
constant & \multicolumn{2}{c}{$2.6\pm 0.4 ~ ~ (\chi^2/ \mathrm{dof} = 1.06)$} \\
\end{tabular}
\end{minipage}
\caption{Determinations of $C_9$ 
from $B\rightarrow K \mmpair$ 
in the low-$q^2$   (left) and
    high-$q^2$   (right) regions. The p-values for the constant fits are 0.17 (low-$q^2$) and 0.39 (high-$q^2$).}
\label{tab:BK}
\end{table}
%%%%%%%%%%%%%%%%%%%%%%%%%%%%%%%%%%%%%%%%%%%%%%%%%%%%%%%%%%%%%%%%%%%%%%%%%%%%%%%%

%%%%%%%%%%%%%%%%%%%%%%%%%%%%%%%%%%%%%%%%%%%%%%%%%%%%%%%%%%%%%%%%%%%%%%%%%%%%%%%%
\begin{table}[t]
\begin{minipage}{0.45\textwidth}
\centering
\begin{tabular}{ c | c | c| c }
$q^2$ (GeV${^2})$ & $C_9^{\parallel}$ & $C_9^{\perp}$& $C_9^{0} $ \\
\midrule
\midrule
$[1.1, 2.5]$ & $2.2_{-1.2}^{+1.3}$ & $6.4_{-1.8}^{+1.7}$ & $1.4_{-0.9}^{+0.9}$\\[2mm]

$[2.5, 4]$ & $4.6_{-1.4}^{+1.4}$ & $3.6_{-1.2}^{+1.3}$ & $2.6_{-1.0}^{+0.8}$ \\[2mm]

$[4, 6]$ & $3.5_{-1.1}^{+1.0} $& $3.5_{-1.0}^{+1.1}$ & $2.4_{-1.2}^{+0.8}$  \\[2mm]

$[6, 8]$ & $3.4_{-0.6}^{+0.6}$ & $2.5_{-0.6}^{+0.6}$ & $3.1_{-0.6}^{+0.6}$\\[2mm]
\midrule
constant & \multicolumn{3}{c}{
$2.8_{-0.2}^{+0.2}$~ $(\chi^2/\mathrm{dof} = 1.26)$} \\
\end{tabular}
\end{minipage}
\hfill
\begin{minipage}{0.45\textwidth}
\centering
\begin{tabular}{ c | c | c| c }
$q^2$ (GeV${^2})$ & $C_9^{\parallel}$ & $C_9^{\perp}$& $C_9^{0} $ \\
\midrule
\midrule
$[11, 12.5]$ & $3.3_{-0.6}^{+0.6}$ & $3.1_{-0.4}^{+0.4}$ & $2.9_{-0.9}^{+0.8}$\\[2mm]

$[15, 17]$ & $3.7_{-0.7}^{+0.6}$ & $3.7_{-0.5}^{+0.5}$ & $3.6_{-0.7}^{+0.7}$ \\[2mm]

$[17,19]$ & $3.4_{-1.0}^{+0.7}$& $4.0_{-0.8}^{+0.8}$ & $3.7_{-0.8}^{+0.8}$  \\[9mm]
\midrule
constant & \multicolumn{3}{c}{
$3.3_{-0.2}^{+0.3} $~ $(\chi^2/\mathrm{dof} = 0.82)$} \\
\end{tabular}
\end{minipage}
\caption{Determinations of $C_9$ in different $q^2$ bins
    from the different polarizations of the $B\rightarrow K^* \mmpair$ decay. The p-values for the constant fits are 0.14 (low-$q^2$) and 0.73 (high-$q^2$).}
\label{tab:BKstar}
\end{table}
%%%%%%%%%%%%%%%%%%%%%%%%%%%%%%%%%%%%%%%%%%%%%%%%%%%%%%%%%%%%%%%%%%%%%%%%%%%%%%%%

We construct the usual $\chi^2$ function as 
\begin{equation}
    \chi^2 = \sum_{i,j} [\mathcal{O}_i^\mathrm{exp}-\mathcal{O}_i^\mathrm{theory}](V_\mathrm{theory}+V_\mathrm{exp})_{ij}^{-1} [\mathcal{O}_j^\mathrm{exp}-\mathcal{O}_j^\mathrm{theory}]
\end{equation}
where the indices $i\,,j$ run over all the observables $\mathcal{O}_{i(j)}$. The matrices $V_\mathrm{theory}$ and $V_\mathrm{exp}$ are the theoretical and experimental covariances, respectively. 
The theoretical covariance is built propagating errors on local form factors, Breit-Wigner parameters, and on $C^\mathrm{eff}_7$. In principle, it has a non-trivial dependence on $C_9^\lambda$. In Ref.~\cite{Altmannshofer:2021qrr}, a method to include such effects in the calculation of the theory covariance matrix is discussed. However, since at the current status experimental uncertainties dominate over the theory one, we choose to set $C_9^\lambda$ to its SM value in the calculation of the theory covariance, effectively accounting for the SM covariance only.
The experimental covariance consists of two parts: the statistical covariance, given in the experimental papers, and the systematic covariance, which we construct from the systematic uncertainties assigning a 100\% correlation. We checked explicitly that this hypothesis does not impact significantly our results, since systematic uncertainties are typically subdominant compared to the statistical ones. Following a frequentist approach, we extract the best-fit point by minimizing the $\chi^2$ function. All the errors correspond to $68\%$ confidence interval, which we obtain by profiling the $\chi^2$ functions over the various fit parameters. We consider two fitting regions: the low $q^2$ region, where $q^2 \in [1.1,8]\,\mathrm{GeV}^2$, and the high $q^2$ region, where $q^2 > 15\, \mathrm{GeV}^2$. In the $B\to K^*\mmpair$ we also consider the bin between the $J/\psi$ and the $\psi(2S)$, $q^2 \in [11,12.5]\,\mathrm{GeV}^2$.

%%%%%%%%%%%%%%%%%%%%%%%%%%%%%%%%%%%%%%%%%%%%%%%%%%%%%%%%%%%%%%%%%%%%%%%%%%%%%%%%
\begin{figure}[t]
    \centering
    \includegraphics[width=0.83\textwidth]{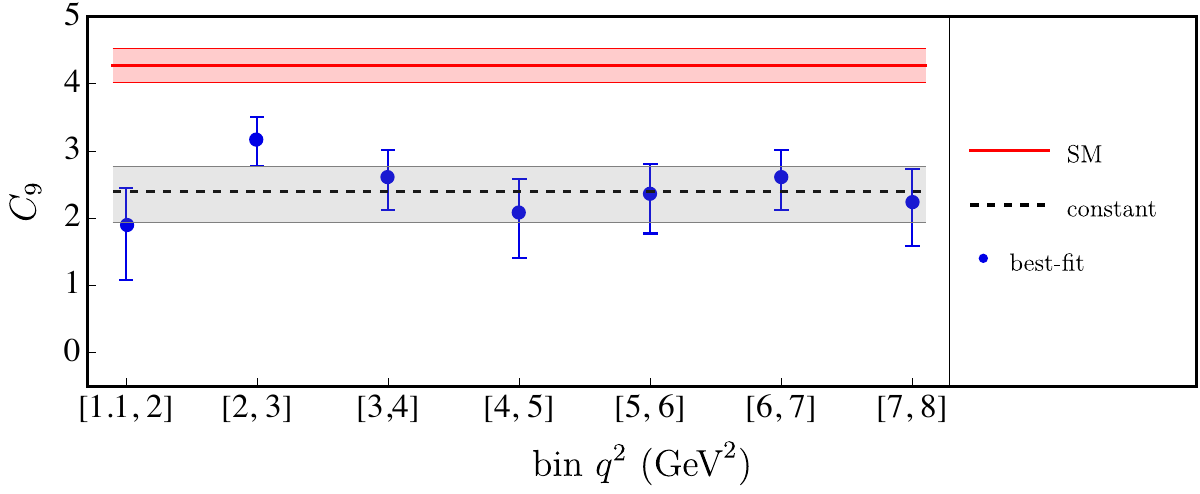}
    \includegraphics[width=0.83\textwidth]{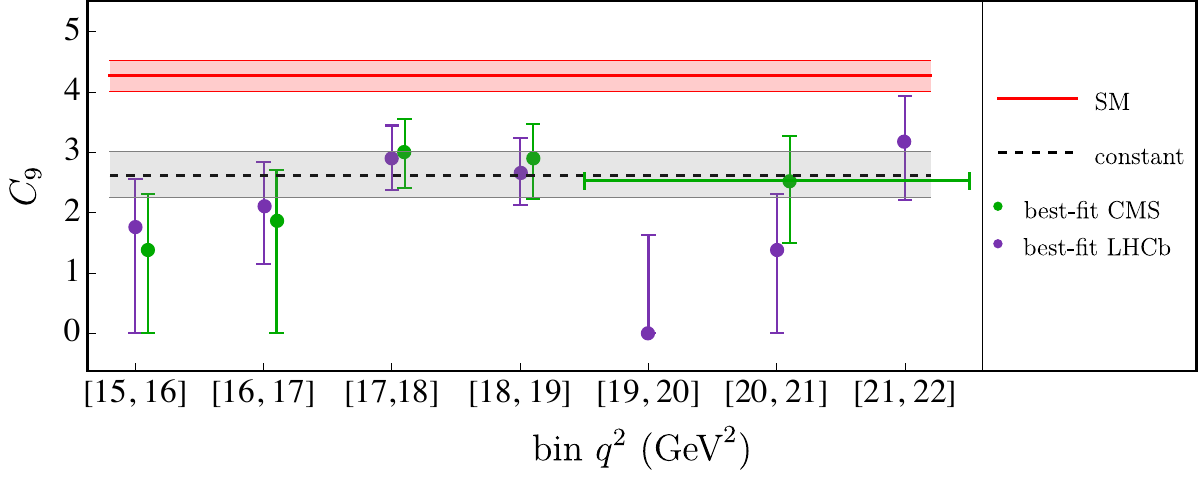}    
    \caption{Determinations of $C_9$ in different $q^2$ bins 
    from $B \rightarrow K \mmpair$ data. The 
    error bars indicate the $68\%$ confidence interval. 
    The red band denotes the SM value. The gray band corresponds to the value extracted assuming a constant ($q^2$-independent) $C_9$, respectively. In the first plot the experimental results from LHCb and CMS have been combined, whereas in the second plot two separate fits are performed  due to different bin widths in the experimental datasets. The bins shown in the figure are the LHCb ones; the best-fit points from CMS data correspond to the bins $(14.82, 16), (16, 17), (17, 18), (18, 19.24), (19.24, 22.9)$ (all in $\mathrm{GeV}^2$).
    } 
    \label{fig:BK}
\end{figure}
%%%%%%%%%%%%%%%%%%%%%%%%%%%%%%%%%%%%%%%%%%%%%%%%%%%%%%%%%%%%%%%%%%%%%%%%%%%%%%%%

For the $B^+\to K^+ \mmpair$ mode, we construct the theory predictions starting from the results of Ref.~\cite{Parrott_2023} and we employ the available data from the LHCb \cite{2014LHCb} and CMS \cite{CMS-PAS-BPH-22-005} collaborations on the differential branching fraction.
The results of the fit are shown in \Table{tab:BK} and Figure~\ref{fig:BK}, 
where we first extract $C_9^K$ in each bin, and then we explore the hypothesis of a constant $C_9^K$ throughout all the kinematic regions. Both at low and high $q^2$, we find that both these hypotheses lead to a good fit, characterized by a $\chi^2/\mathrm{dof}$ close to unity.

\begin{figure}[t]
    \centering
    \includegraphics[width=0.83\textwidth]{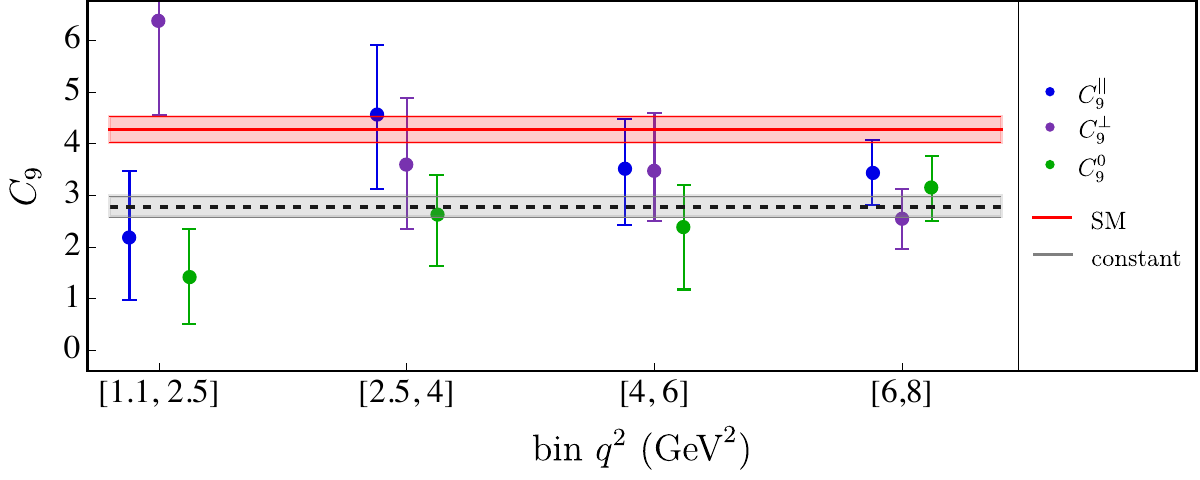}
    \includegraphics[width=0.83\textwidth]{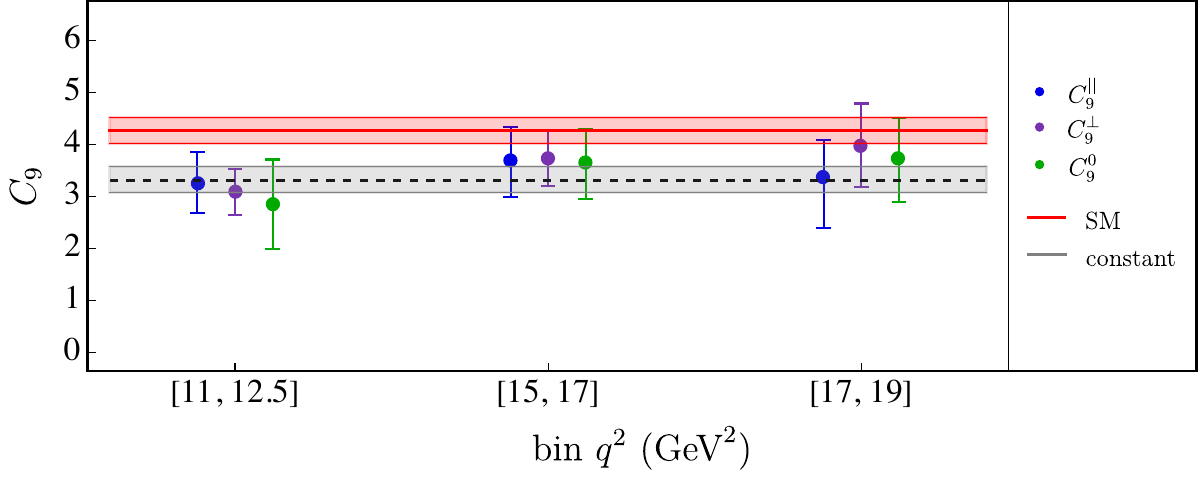}
    \caption{Determinations of $C_9$ in different $q^2$ bins
    from the different polarizations of the $B\rightarrow K^* \mmpair$ decay.
    Notations as in Figure~\ref{fig:BK}.} 
    \label{fig:BKstar}
\end{figure}

In the case of $B \rightarrow K^{*0} \mmpair$ we fit the branching ratio  using the experimental results reported in \cite{LHCb:2016ykl}, as well as the angular observables $F_L, S_3, S_4, S_5, A_{FB},S_7, S_8, S_9$, measured by LHCb in \cite{LHCb:2020lmf}. We implement the angular observables as in \cite{Altmannshofer:2008dz,Bobeth:2012vn} and we correct for the mismatch in the definition of the muon helicity angle to follow the experimental conventions given in \cite{LHCb:2013ghj}.
To obtain the theory predictions we use the form factors from \cite{Straub:2015ica}; see also \cite{Gubernari:2023puw} for a recent review. Due to the lack of data on the decays $B \rightarrow K^* V_j$ with $V_j$ above the $\psi(2S)$ resonance, only the first two $c \bar{c}$ resonances, the $J/\psi$ and the $\psi(2S)$, are implemented.
The results of the fits are reported in Table~\ref{tab:BKstar} and   Figure~\ref{fig:BKstar}. At the bottom of Table~\ref{tab:BKstar} the results of the fit in the low-$q^2$ and in the high-$q^2$ regions under the assumption of a constant $C_9$, averaged over the bins and the different polarizations, 
are also indicated.

%%%%%%%%%%%%%%%%%%%%%%%%%%%%%%%%%%%%%%%%%%%%%%%%%%%%%%%%%%%%%%%%%%%%%%%%%%%%%%%%
%%%%%%%%%%%%%%%%%%%%%%%%%%%%%%%%%%%%%%%%%%%%%%%%%%%%%%%%%%%%%%%%%%%%%%%%%%%%%%%%

\section{Discussion}
\label{sect:disc}

As discussed in Section~\ref{sect:theory},
treating  $C_9$  as a $q^2$- and
channel-dependent quantity allows us to describe in full generality the 
long-distance contributions to the $B \rightarrow K^{(*)}  \mu^+ \mu^-$ amplitudes
of SM origin. 
In the limit where the $Y$ functions 
in Eq.~(\ref{eq:C9fit}) describe well these long-distance effects,
we should expect the experimentally determined 
$C^\lambda_9 (q^2)$ values to exhibit no
$q^2$ and $\lambda$ dependence (within errors).
Moreover, the extracted values should coincide with the SM prediction of  $C_9(\mu_b)$.
Conversely,  a dependence from $q^2$ 
and/or  $\lambda$ in the   values 
of $C^\lambda_9 (q^2)$ thus determined would unambiguously signal an incorrect description of long-distance dynamics via the $Y$ functions.

The independent $C^\lambda_9 (q^2)$ values determined from data, 
reported in Tables~\ref{tab:BK}--\ref{tab:BKstar}, exhibit no 
significant $q^2$  and/or  $\lambda$ dependence. This statement
is  evident if we look at Figures~\ref{fig:BK}--\ref{fig:BKstar}, 
where the results for the low- and high-$q^2$
regions are shown separately for the two modes. 
However, it also holds in the whole $q^2$ spectrum and for all the hadronic amplitudes.
To better quantify this statement, in Table~\ref{tab:common} we report the 
results of fits performed assuming constant $C_9$ values in the low- and high-$q^2$ regions,
separating or combining the different decay amplitudes, or considering the same 
value over the full spectrum for all the decay amplitudes.

\begin{table}[p]
        \centering
        \renewcommand{\arraystretch}{1.5}
\begin{tabular}{c|l||c|c|c|c}
    $q^2$ region & Amplitude & \multicolumn{4}{c}{$C_9$ values }  \\
    \hline \hline
    \multirow{4}{*}{Low $q^2$}&$B\rightarrow K$ & \multicolumn{2}{c|}{$2.4_{-0.5}^{+0.4}$} & \multirow{4}{*}{$2.7_{-0.2}^{+0.2}$} &   \\
    \cline{2-4}
    &$ B\rightarrow K^*(\epsilon_{\parallel})$ & $3.0_{-0.6}^{+0.6}$ & & \\
    \cline{2-3}
    &$ B\rightarrow K^*(\epsilon_{\perp})$ & $2.7_{-0.7}^{+0.7}$ &  $2.8_{-0.2}^{+0.2}$ & ${}_{(\chi^2/\mathrm{dof} = 1.28,~\text{p-value=0.09)}}$ \\
    \cline{2-3}
    &$ B\rightarrow K^*(\epsilon_{0})$ & $2.7_{-0.8}^{+0.7} $&  & &  $3.0_{-0.1}^{+0.1}$  \\
    \cline{1-5}
    \multirow{4}{*}{High $q^2$}&$B\rightarrow K$ & \multicolumn{2}{c|}{$2.6_{-0.4}^{+0.4}$} & \multirow{4}{*}{$3.0_{-0.2}^{+0.2}$} & $(\chi^2/\mathrm{dof} = 1.33,~\text{p-value=0.01})$  \\
    \cline{2-4}
    &$ B\rightarrow K^*(\epsilon_{\parallel})$ & $3.2_{-0.5}^{+0.5}$ & &  \\
    \cline{2-3}
    &$ B\rightarrow K^*(\epsilon_{\perp})$ & $3.4_{-0.4}^{+0.4}$ &  $3.3_{-0.2}^{+0.3}$ & ${}_{(\chi^2/\mathrm{dof} = 1.06,~\text{p-value}=0.37)}$ \\
    \cline{2-3} 
    &$ B\rightarrow K^*(\epsilon_{0})$ & $3.3_{-0.6}^{+0.6}$ &  &  \\
    \hline
\end{tabular}
\caption{Best-fit points assuming constant $C_9$ values in the low- and high-$q^2$ regions, separating or combining the different decay amplitudes, or considering the same value  over the full $q^2$ spectrum for all the decay amplitudes (last column). }
\label{tab:common}
\end{table}
\begin{figure}[p]
    \centering
    \includegraphics[scale=0.7]{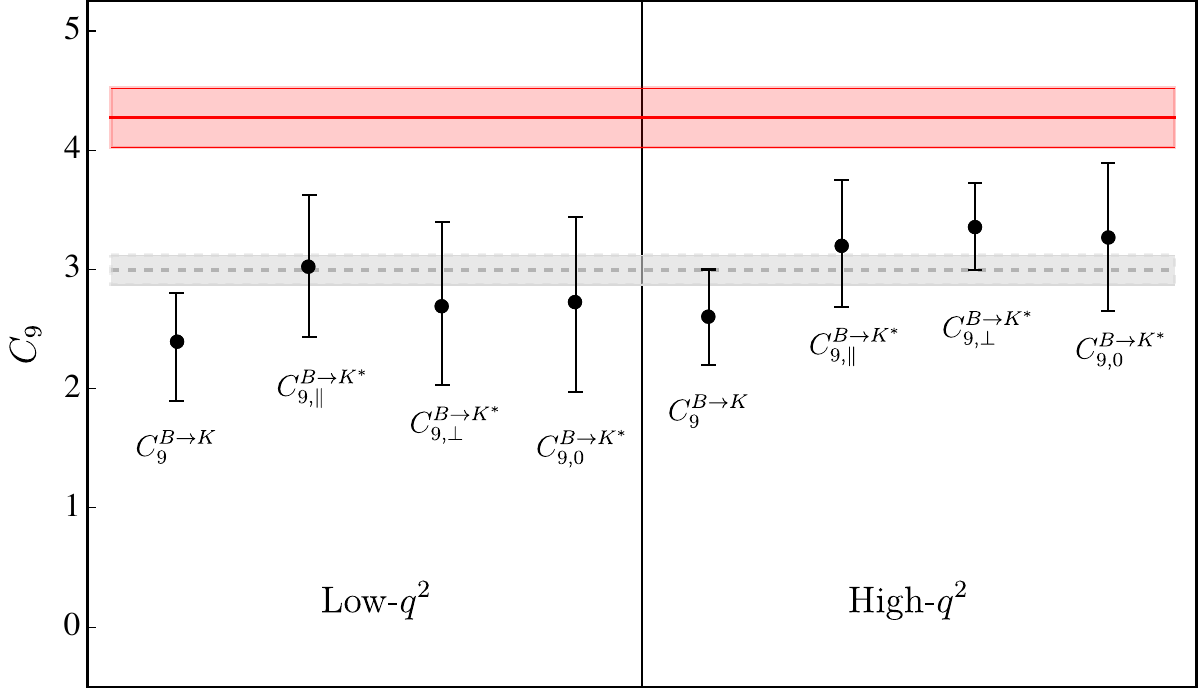}
    \caption{Independent determinations of $C_9$. The black points illustrate the determinations 
    in the low- and high-$q^2$ regions for the different decay amplitudes (see Table~\ref{tab:common}).
    The grey band is the result of the fit assuming a universal $C_9$ over the full spectrum. The red 
     band indicates the SM value.  }
    \label{fig:common}
\end{figure}

A graphical illustration of the combined 
fit results is shown in Figure~\ref{fig:common}.
As can be seen, the independent $C_9$ values determined in different kinematical regions and in
different hadronic amplitudes are all well consistent. A  
quantification of the  consistency is provided by the $\chi^2$ 
of the fit assuming a universal $C_9$, namely $\chi^2/\mathrm{dof}=1.33$.
On the other hand, it is evident that the 
universal  $C_9$ determined from data is not in good agreement with the SM expectation. 
The two main conclusions we can derive from these
results can be summarized as follows:
\begin{itemize}
    \item Data provide no evidence of sizable unaccounted-for long-distance contributions. These would naturally lead to a significant  $q^2$ and/or $\lambda$ dependence in the experimentally determined 
    $C^\lambda_9 (q^2)$ values, that we do not observe (at least given the present level of precision). 
    \item The observed discrepancy in the experimentally determined 
    $C_9$ value, compared to the SM expectation, is consistent with 
    a short-distance effect of non-SM origin.  
\end{itemize}
These findings are qualitatively similar to those obtained in Ref.~\cite{Alguero:2023jeh}.

In Table~\ref{fig:common} we also report the p-value of the fits performed under the hypothesis of a constant $C_9$. The overall p-value of $1\%$, when all data are combined, indicates that the global fit is not particularly good.
The main source of this low probability is the precise $B\to K \mu^+\mu^-$ data at low-$q^2$ (Table~\ref{fig:BK}, upper panel). Indeed, excluding these data, the p-value raises to  $15\%$. The origin of this discrepancy is not clear at this stage (could be a problem in the experimental data, in the  $B\to K$ form factors at low $q^2$, or a sub-leading re-scattering effect not well described). However, given the fitted value of $C_9$ from $B\to K$ at low $q^2$ is in good agreement with the other seven determinations, the low p-value of those data does not modify the main conclusion of a largely $q^2$-- and $\lambda$-- independent shift in $C_9$ compared to the SM expectation.

In principle, we cannot exclude a sizable long-distance contribution
 with no $q^2$ and  $\lambda$ dependence, which would mimic a short-distance effect.
 However, this is a rather unlikely possibility. To this purpose, it is worth noting that the known long-distance contributions, described by the $Y$ functions, exhibit a strong $q^2$ and  $\lambda$ dependence. In particular, an estimate of the $\lambda$ dependence can be obtained by comparing the different $\eta^\lambda_V$ values 
 (for a given resonance) reported in Tables~\ref{tab:EtasPhisBtoK}--\ref{tab:EtasPhisBtoKstar}. They vary up to a factor of $3$ in the $V=J/\Psi$ case and up to a factor of $6$ in the $V=\psi(2S)$ case.

To conclude, we stress that the uncertainties of the independent $C_9$ values reported in  Figure~\ref{fig:common} are still rather large. This partially weakens the two statements above, and is the reason why we refrain from quantifying (in terms of $\sigma$'s) the discrepancy between data and the SM hypothesis. If the absence of $q^2$ and $\lambda$ dependence survived with smaller uncertainties, the implausibility of the hypothesis of unaccounted-for long-distance contributions would emerge more clearly. This, in turn, would enable a credible quantitative estimate of the deviation from the SM hypothesis.

\section{Conclusions}
\label{sect:conc}

The difficulty of performing precise SM tests 
in $B\to K^{(*)} \ell^+\ell^-$ decays lies in the difficulty of 
precisely estimating non-perturbative long-distance 
contributions related to charm re-scattering in these rare modes.
In this paper we have presented a general amplitude decomposition that allows us to describe these effects
in full generality,  in both $B\to K \ell^+\ell^-$ and $B\to K^* \ell^+\ell^-$ decays, and over the full $q^2$ spectrum.

Using this  general amplitude decomposition we have analyzed the 
available  $B\to K^{(*)} \mu^+\mu^-$ data obtaining  
independent determinations of the Wilson coefficient $C_9$ 
from different kinematical regions and from different 
hadronic amplitudes. The results, summarized in 
 Figure~\ref{fig:common},  do not indicate a significant dependence on $q^2$ and/or the hadronic channel, as naturally expected in the case of  unaccounted-for  long-distance contributions.
 On the other hand, they exhibit a  systematic shift compared to the SM value. 
 These findings support the hypothesis of a non-standard $b\to s \mmpair$ amplitude of short-distance origin.

At present, given the sizable uncertainties of the independent $C_9$ values reported in  Figure~\ref{fig:common}, and the relatively low p-value
of the global fit obtained assuming a universal $C_9$, is difficult to translate the above qualitative conclusion into a quantitative statement about the inconsistency of the SM hypothesis.  However, the method we have presented has no intrinsic theoretical limitations: with the help of more precise data and more precise determinations of the local form factors from Lattice QCD, it could allow us to derive more precise results. 
A crucial ingredient is also the determination of the resonance parameter directly from data, which is presently available only in the $B\to K$ case. 
 If, with the help of more data, the absence of $q^2$ and $\lambda$ dependence survived with smaller uncertainties,  the implausibility of the hypothesis of unaccounted-for long-distance contributions might emerge more clearly. This, in turn, would enable a reliable quantitative estimate of the deviation from the SM hypothesis.

\section*{Acknowledgments}
We thank M\'eril Reboud and Peter Stangl for useful discussions.
This project has received funding from the European Research Council~(ERC) under the European Union's Horizon~2020 research and innovation programme under grant agreement 833280~(FLAY), and by the Swiss National Science Foundation~(SNF) under contract~200020\_204428.
 
\appendix

\section{Resonance parameters}
\label{app:etaV}

\subsection{$B \rightarrow K \ell^+ \ell^-$   }
        Defining the $\eta^K_V$ parameters as in Eq.~(\ref{eq:etaVdef}),
        the branching ratio for the resonance-mediated process 
        $B^+ \rightarrow K^+ V   \to  K^+ \mu^+ \mu^-$ is 
        \begin{align}\label{eq:BRJpsi}
            &\mathcal{B}\left(B^+ \rightarrow K^+ V  \to  K^+ \mu^+ \mu^- \right) = \left|\eta^K_{V} \right|^2 \frac{G_\mathrm{F}^2 \alpha_\mathrm{em}^2 \left\vert V_{cb} V_{cs}^* \right\vert^2}{1024 \pi^5 \Gamma_{B^+}} \times 
            \nonumber \\
           & \times \int_{4m_\mu^2}^{\left(m_B - m_K \right)^2} \left\vert \mathbf{k}(q^2) \right\vert^3 \left[\beta(q^2) - \frac{1}{3}\beta^3(q^2) \right] \left(16 \pi^2 \right)^2 \left|A_{V }^\mathrm{res} \left(q^2 \right) \right|^2 \left|\frac{q^2}{m_{V }^2} \right|^2 \mathrm{d}q^2 ,
        \end{align}
        where  $\beta(q^2) = \sqrt{1-4 m_\mu^2/q^2}$ and $\left|\mathbf{k} (q^2)\right| = \lambda^{1/2}(m_B^2,m_K^2,q^2)/ 2m_B $.
        In the narrow-width approximation, which is an excellent 
        approximation for the $J/\psi$ and $\psi(2S)$ resonances,  
        we obtain the following expression for $|\eta^K_V|$: 
        \begin{align}
        \left|\eta^K_{V } \right|  &= [\mathcal{B}\left(B^+ \rightarrow K^+ V   \to  K^+ \mu^+ \mu^- \right) ]^{1/2} \times \nonumber \\
          &\times \left[   \frac{  G_\mathrm{F}^2 \alpha_\mathrm{em}^2 \left|V_{cb} V_{cs}^* \right|^2 \Gamma_{V } m_{V } \lambda^{3/2}\left(m_{B^+}^2,m_{K^+}^2, m_{V }^2 \right) }{6 m_{B^+}^3 \Gamma_{B^+}} \right]^{-1/2}\,.
        \end{align}
We have used this expression, together with the $\mathcal{B}\left(B^+ \rightarrow K^+ V   \to  K^+ \mu^+ \mu^- \right)$ reported in~\cite{LHCb:2016due}, to derive the $\eta^K_V$
 reported in Table~\ref{tab:EtasPhisBtoK}. The corresponding 
$\delta^K_V$ have been determined by the LHCb collaboration~\cite{LHCb:2016due}
considering the interference between resonant and non-resonant amplitudes.

\subsection{$B \rightarrow K^* \ell^+ \ell^-$  }
The discussion of the $K^*$ case is a bit more involved. We start from a general decomposition of the $B\to K^* V$ weak matrix element that, following \cite{Dighe:1998vk}, reads
    \begin{equation}     \bra{K^*(p_{K^*},\epsilon(\lambda)) V(q,\eta(\lambda_V))}  \cL_\mathrm{eff}(b\to s)   \ket{B_q(p_B)} = 
        \epsilon_{\mu}^*(\lambda)\eta_{\nu}^*(\lambda_V) M^{\mu\nu}
    \end{equation}
    where
    \begin{equation}
        M^{\mu\nu} = ag^{\mu\nu}+\frac{b}{m_{K^*} m_V}q^\mu p_{K^*}^\nu+i \frac{c}{m_{K^*} m_V}\epsilon^{\mu\nu\alpha\beta}p_{K^*\alpha}q_{\beta}~.
        \label{eq:BtoVectorsDighe}
    \end{equation}
The coefficients $a$, $b$ and $c$  encode all  possible 
contractions of the four-quark charm operators ($\cO_{1,2}$), which provide the largely dominant contribution to the amplitude. 
It is convenient to introduce three helicity amplitudes, which are related to the coefficients $a$, $b$ and $c$ as
    \begin{align}
        \mathcal{A}_0 =&  -x a -(x^2-1) b\,, \nonumber \\
        \mathcal{A}_\parallel=& +\sqrt{2}\, a\,,  \nonumber \\
        \mathcal{A}_\perp=& +\sqrt{2(x^2-1)}\, c\,,
        \label{eq:curlAdef} 
    \end{align}
    where $x = p_{K^*}\cdot p_V/ (m_{K^*} m_V)$.

We further parameterize the matrix element of the electromagnetic current relevant 
to the  $V \to e^{+} e^{-}$ decay as 
    \begin{align}
    \braket{V(q,\eta(\lambda_V))| j^{\mu}_{\rm em}|  0} = 2  f_{V}(q^{2}) \eta^{\mu \ast}(\lambda_V) \,.
    \label{eq:fVAdef} 
    \end{align}
Here  $f_{V}\equiv f_V(m_V^2)$ is a dimensionless constant that can be determined by $\mathcal{B}(V \to \ell^{+} \ell^{-})$ via the relation
    \begin{align}
     \mathcal{B}(V \to \ell^{+} \ell^{-}) = \frac{4\alpha_{\mathrm{em}} f_V^2(m_V^2) m_V}{3 \Gamma_V} \,.
    \end{align}
    In the cases we are interested in, namely $V= J/\psi$ and $\psi(2S)$, 
    using the leptonic branching fractions for the charmonium states in \cite{Zyla:2020zbs}, we derive
    \begin{equation}
    f_{J/\psi}=(1.36\pm 0.02)\times 10^{-2}\,, \qquad
    f_{\psi(2S)} = \left(4.87 \pm 0.07 \right) \times 10^{-3}\,.
    \end{equation}

 Combining Eqs.~(\ref{eq:BtoVectorsDighe}), (\ref{eq:curlAdef}) and (\ref{eq:fVAdef}) we obtain
        \begin{align}
          & \mathcal{M}(B\to K^* V \to K^*\ell^+\ell^-)= \sum_{\lambda_V} \mathcal{M}\left(B\rightarrow K^* V \right) \mathcal{M}\left(V \rightarrow \ell^+ \ell^- \right)  \nonumber \\
            =&\frac{2 e f_{V}}{q^2-m_{V}^2+im_{V}\Gamma_{V}} \Bigg[\frac{\mathcal{A}_\parallel}{\sqrt{2}}\left(-\epsilon^{\ast\rho} (\lambda) + \frac{\epsilon^\ast (\lambda) \cdot q}{m_{V}^2 m_{K^*}(x^2-1)} (x m_V p_{K^*}^\rho-m_{K^*}q^\rho) \right)  \nonumber \\
            &\phantom{\frac{2 e f_{V}}{q^2-m_{V}^2-im_{V}\Gamma_{V}} \Bigg[} + \frac{\mathcal{A}_0}{m_{K^*}m_V^2(x^2-1)}\left(\epsilon^\ast (\lambda)\cdot q \right)\left( m_V p_{K^*}^\rho-m_{K^*}x q^\rho \right) \nonumber \\
            &\phantom{\frac{2 e f_{V}}{q^2-m_{V}^2-im_{V}\Gamma_{V}} \Bigg[} -\frac{i\mathcal{A}_\perp}{\sqrt{2\left(x^2 -1 \right)} m_V m_{K^*}} \epsilon^{\mu\alpha\beta\rho} \epsilon_{\mu}^\ast (\lambda) p_{K^*\alpha} q_\beta \Bigg] \overline{\ell}\gamma_\rho \ell
            \label{eq:BToResonanceToLeptonsME}
        \end{align}
By equating  \eq{eq:BToResonanceToLeptonsME} to (\ref{eq:BToHII0}),
taking into account the definition of the 
$\cH_{\rm{c\bar c}}^\lambda(q^2)$ functions  in (\ref{eq:BKsdec}) 
and that of the 
$\{\eta_V^\lambda,\delta_V^\lambda\}$ parameters in  (\ref{eq:etaVdef}), 
we deduce 
        \begin{align}
            \begin{aligned}
                \eta_{V}^0 e^{i\delta_{V}^0} &= - \frac{f_{V}}{8\sqrt{2\pi \alpha} G_\mathrm{F} |V_{tb} V^*_{ts} | m_B m_{K^*} \Gamma_{V}} \mathcal{A}_0\,, \\
                \eta_{V}^\parallel e^{i\delta_{V}^\parallel} &=- \frac{f_{V}}{2\sqrt{\pi\alpha} G_\mathrm{F} |V_{tb} V_{ts}^*|  (m_B+m_{K^*}) m_{V} \Gamma_{V}} \mathcal{A}_\parallel\,, \\
                \eta_{V}^\perp e^{i\delta_{V}^\perp} &=- \frac{f_{V} (m_B + m_{K^*})}{2\sqrt{\pi\alpha\lambda} G_\mathrm{F} |V_{tb} V_{ts}^*|  m_{V}\Gamma_{V}} \mathcal{A}_\perp\,.
            \end{aligned} \label{eq:EtaPhi}
        \end{align}
        The negative signs in \eq{eq:EtaPhi} take into account 
        the negative sign of Re($V_{ts}$) in the standard CKM phase convention, that we adopt through the paper.

 The last step in order to derive numerical predictions for the 
 $\{\eta_V^\lambda,\delta_V^\lambda\}$ is to extract magnitudes and phases of the 
 $\cA_\lambda$ taking into account experimental data on decay rates and time-dependent distributions of $B\to K^* V$ decays. 
 Experimentally, $B\to K^* V$ decays are analyzed in terms of the 
  normalized helicity amplitudes $A_\lambda(t)$, 
     \begin{equation}
        \frac{d\Gamma(B\to K^* V)}{dt} = \mathcal{N} (| A_0(t)|^2+|A_\parallel(t)|^2+|A_\perp(t)|^2) \,,
        \label{eq:time_dependent_rate}
    \end{equation}
satisfying 
\begin{equation}
 |A_0(0)|^2+ |A_\parallel(0)|^2+ |A_\perp(0)|^2=1\,.
 \label{eq:Aunit}
\end{equation}
The explicit time dependence of the $A_\lambda(t)$ 
can be found in \cite{LHCb:2013vga}.
By comparing the time integral of \eq{eq:time_dependent_rate}, 
with the partial rate expressed in terms of the $\cA_\lambda$, 
namely
    \begin{equation}
        \Gamma(B\to K^* V) = (| \mathcal{A}_0|^2+|\mathcal{A}_\parallel|^2+|\mathcal{A}_\perp|^2)\frac{ \lambda^{1/2} (m_B^2, m_V^2, m_{K^*}^2) }{16\pi m_B^3}\,,
    \end{equation} 
    we deduce 
    \begin{align}
        |\mathcal{A}_i|^2 =&\, (2.46\times 10^{-13} \mathrm{GeV}^2) \times |A_i(0)|^2  \,.
    \end{align}
in the $B\to K^* J/\psi$ case.
  
    The complex amplitudes $A_i(0)$ are  parameterized as 
    $A_i(0) = |A_i(0)| e^{-i\delta_i}$
    and, by convention, data are analyzed setting $\delta_0 = 0$. 
    The  experimental values thus determined 
    in the $B\to K^* J/\psi$ case are~\cite{LHCb:2013vga} 
    \begin{equation}\label{eq:MeasuredAmpsBKstar}
    \begin{aligned}
        |A_0(0)|^2  =&\, 0.227\pm 0.004\pm0.011\,, & \quad \delta_\parallel =&\, -2.94\pm0.02\pm0.03\,,  \\
        |A_\perp(0)|^2  =&\, 0.201\pm 0.004\pm0.008\,, & \quad \delta_\perp =&\, 2.94\pm0.02\pm0.02\,,
    \end{aligned}
    \end{equation}
    with $|A_\parallel(0)|$ unambiguously fixed by \eq{eq:Aunit}.
    In the absence of experimental data on $B\to K^* J/\psi (\to \mu^+ \mu^-)$ decays fixing the overall phase difference between resonant and non-resonant amplitudes, we assume this interference to be the same as in the 
    $B\to K J/\psi$ case~\cite{LHCb:2016due}, where the two amplitudes are almost orthogonal in the complex plane: $\phi_{\rm rel} \approx -1.5$.
    We thus shift all the $\delta_\lambda$ in 
    \eq{eq:MeasuredAmpsBKstar}, as well as $\delta_0$, by 
    $\phi_{\rm rel}$.    
    Using these results in \eq{eq:EtaPhi} leads to the 
     $\eta_{J/\psi}^\lambda$ and $\delta_{J/\psi}^\lambda$ values 
     reported in Table~\ref{tab:EtasPhisBtoKstar}.  
This assumption about the overall phase difference is certainly very  na\"ive; however, we have checked explicitly that varying this phase 
by $\pm30\%$ has a negligible numerical impact on the analysis.

    \subsubsection*{Determining $B\to K^* \psi(2S)$ amplitudes  
   via  \texorpdfstring{$SU(3)_\mathrm{F}$}{} relations}

Since the $B\to K^* \psi(2S)$ helicity amplitudes have not been measured, we use $\mathrm{SU}(3)_F$ relations to estimate them in terms of the 
$B_s \rightarrow \psi(2S) \phi$ ones, which are experimentally accessible. 
In the $B_s$ case, the $t=0$ normalized helicity amplitudes for the 
two narrow resonances reads~\cite{LHCb:2016tuh} 
        \begin{equation}\label{eq:MeasuredAmpsBsPhiPsi}
        \begin{aligned}
            \left\vert A_\perp (B_s \rightarrow J/\psi \phi) \right\vert^2  =&\, 0.2504\pm 0.0049\pm 0.0036\,, & \quad \delta_\perp =&\, 
            3.08^{+0.14}_{-0.15}\pm0.06\,,\\
            \left\vert A_0  (B_s \rightarrow J/\psi \phi) \right\vert^2  =&\,  
            0.5241 \pm 0.0034 \pm 0.0067\,, &  \quad \delta_\parallel =&\, 3.26^{+0.10}_{-0.17}\pm0.06\,, \\
        \end{aligned}
        \end{equation}
        \begin{equation}\label{eq:MeasuredAmpsBsPhiPsi2S}
        \begin{aligned}
            \left\vert A_\perp (B_s \rightarrow \psi(2S)\phi) \right\vert^2  =&\, 0.264_{-0.023}^{+0.024}\pm 0.002\,, & \quad \delta_\perp =&\, 3.29_{-0.39}^{+0.43}\pm 0.04\,, \\
            \left\vert A_0  (B_s \rightarrow \psi(2S)\phi) \right\vert^2  =&\, 0.422\pm0.014 \pm 0.003\,, & \quad \delta_\parallel =&\, 3.67^{+0.13}_{-0.18}\pm0.03\,,  \\
        \end{aligned}
        \end{equation}        
        yielding the ratios     
        \begin{equation}
            \begin{aligned}
                \left| \frac{A_0\left(B_s \rightarrow \psi(2S) \phi \right)}{A_0 \left(B_s \rightarrow J/\psi \phi \right)} \right| &= 0.897 \pm 0.017\,, &  \\
                \left| \frac{A_\perp\left(B_s \rightarrow \psi(2S) \phi \right)}{A_\perp \left(B_s \rightarrow J/\psi \phi \right)} \right| &= 1.03 \pm 0.33, & \frac{\delta_\perp\left(B_s \rightarrow \psi(2S) \phi \right)}{\delta_\perp \left(B_s \rightarrow J/\psi \phi \right)} &= 1.08 \pm 0.14\,, \\
                \left| \frac{A_\parallel\left(B_s \rightarrow \psi(2S) \phi \right)}{A_\parallel \left(B_s \rightarrow J/\psi \phi \right)} \right| &= 1.18 \pm 0.32, & \frac{\delta_\parallel \left(B_s \rightarrow \psi(2S) \phi \right)}{\delta_\parallel \left(B_s \rightarrow J/\psi \phi \right)} &= 1.13 \pm 0.07\,.
            \end{aligned}
        \end{equation}        
        Assuming the same ratios hold in the 
        $B \rightarrow K^\ast V$ case, 
        we determine the corresponding $B\to 
        K^* \psi(2S)$ amplitudes starting from the $B\to 
        K^* \psi$ ones in Eq.~(\ref{eq:MeasuredAmpsBKstar}).
        The phases are then corrected for an overall shift that we deduce from the $B\to K \psi(2S)$ result.
        Also in this case we  have checked that varying this phase difference  
by $\pm30\%$ has a negligible impact on the analysis.

\bibliographystyle{utphys}
\bibliography{references}
\end{document}